\documentclass[review]{elsarticle}
\usepackage{graphicx}
\usepackage{amsmath,bm}
\usepackage{booktabs}
\usepackage{threeparttable}
\usepackage{indentfirst}
\usepackage{url}
\usepackage{lineno,hyperref}
\usepackage{lineno,hyperref}
\modulolinenumbers[5]
\begin{document}

\begin{frontmatter}

\title{Recognition of Emotions using Kinects}

% Group authors per affiliation:
\author{Shun Li$^1$$^,$$^2$,Changye Zhu$^3$,Liqing Cui$^1$$^,$$^2$,Nan Zhao$^1$,Baobin Li$^3$,Tingshao Zhu$^1$}
\address{$^1$Chinese Academy of Sciences Institute of Psychology,$^2$The 6th Research Institute of China Electronics Corporation,$^3$University of Chinese Academy of Sciences}

%%$\ead[url]{www.elsevier.com}

\begin{abstract}
\indent
Psychological studies indicate that emotional states are expressed in the way people walk and the human gait is investigated in terms of its ability to reveal a person's emotional state. And Microsoft Kinect is a rapidly developing, inexpensive, portable and no-marker motion capture system. This paper gives a new referable method to do emotion recognition, by using Microsoft Kinect to do gait pattern analysis, which has not been reported. $59$ subjects are recruited in this study and their gait patterns are record by two Kinect cameras. Significant joints selecting, Coordinate system transforming, Slider window gauss filter, Differential operation, and Data segmentation are used in data preprocessing. Feature extracting is based on Fourier transformation. By using the NaiveBayes, RandomForests, libSVM and SMO classification, the recognition rate of natural and unnatural emotions can reach above $70\%$.It is concluded that using the Kinect system can be a new method in recognition of emotions.
\end{abstract}

\begin{keyword}
\texttt{emotion recognition,Microsoft Kinect,gait patterns}%\sep \LaTeX\sep Elsevier \sep template
%\MSC[2010] 00-01\sep  99-00
\end{keyword}

\end{frontmatter}

%\linenumbers

\section{Introduction}

\indent
Detection of emotions is generally based on observing facial expressions, linguistic as well as acoustic feature in speech, physiological parameters, gesture and body motions\cite{bealeaffect}.Body motions are considered as additional modality to estimate the emotion state of a human. Affective body movement provides important visual cues used to distinguish expression of emotion\cite{pollick2001perceiving}.Since gait is a natural day to day motion, our work concentrates on the recognition of emotions in gait patterns.\\
\indent
Psychological studies indicate that emotional states are expressed in the way people walk and the human gait is investigated in terms of its ability to reveal a person¡¯s emotional state. In recent years, much research has been done to define a normal gait patterns. This task is challenging, because people¡¯s individual gait is as unique as their fingerprint\cite{1323098}.Furthermore, gait is influenced by many factors such as age, weight, and possible gait disorders. How such factors or combinations of them affect gait is still an open question.\\
\indent Our work gives a new referable method to do emotion recognition, by using Microsoft Kinect to do gait pattern analysis, which has not been reported. The Microsoft Kinect is camera-based sensor primarily used to directly control computer games through body movement. The tracks of the position of the limbs and body without the need for handheld controllers or force platforms. Use of a depth sensor also allows the Kinect to capture three-dimensional movement patterns. The system software enables feature extraction to recognize body joint centers. In comparison to conventional motion capture systems, Kinect system is low-cost\cite{Gaukrodger2013},portable,no- marker\cite{Pogrzeba2013},and easy to use\cite{Stone201557}.\\
\indent
We recruit the subjects and collect the joints¡¯ timing sequence data by using Kinect system, then use machine learning method to do feature extraction and classification, and the recognition rate between the natural and unnatural emotions can reach above $70\%$.\\
\indent
The remainder of this paper is organized as follow. Section 2 introduces the related work of recognition of emotions in walking and the Kinect application. Section 3 describes the method of our work, including the experiment, the database, and the data processing. Classification and results are presented in section 4 and discussed in section 5.The paper ends with a conclusion in section 6.
\section{Related work}
\indent
In psychology, evidence exists that emotion can be expressed in walking and recognized by human observers. In 1987,Montepare et.al's psychological study indicated that observers can identify emotion from variations in walking style\cite{7}.In 2008,Janssen used the conventional camera to acquire kinetic and kinematic data and do recognition of emotions in
gait\cite{8} and in 2009-2010,Krag did Person-Dependent and Inter-Individual Recognition of emotions by marker-based gait analysis using motion capturing system\cite{karg2009two,5349438,5439949}.But conventional 3-dimension video-based motion analysis systems allow for comprehensive kinematic and kinetic analysis of gait and require large spaces, and considerable expertise and are expensive. The marker-based motion capture systems (e.g. The VICON system) require precise, tedious and time-consuming maker preparation, which may affect the subjects¡¯ emotional states and also expensive.\\
\indent
Low cost options could include inertial monitoring sensors such as accelerometers\cite{Wang20111780} and gyroscopes, however these sensors possess sources of error such as signal drift and noise which impede their accuracy\cite{Clark20131064}.\\
\indent
Microsoft Kinect is a rapidly developing, inexpensive, portable and no-marker motion capture system. Early reports suggests the Kinect can identify pose\cite{6096866} and simple stepping movements\cite{6338001} in healthy adults. Recently, clinical researchers have reported interesting applications using the Kinect system. for example, an interactive game-based rehabilitation tool for balance training\cite{6090521} and and a 3-D body scanner\cite{6165146}.And methods have been proposed to detect gait patterns in walking data obtained with Kinect: In 2014,Auvinet used a Kinect to detect the gait cycles in treadmill\cite{Auvinet2015722};Yeung evaluated the Kinect as a clinical assessment tool of body sway\cite{Yeung2014532};Galna used Kinect to measure the movement in people with Parkinson¡¯s disease\cite{Galna20141062}.\\
\indent
Previous studies have validated the Kinect as a motion capture system. Accuracy and sensitivity of kinematic measurements obtained from Kinect, such as reaching distance, joint angles, and spatial-temporal gait parameters, were comparable to a VICON system\cite{khoshelham2012accuracy}.Evidence exists that it can accurately assess the gait patterns dynamics during walking\cite{stone2011evaluation}.\\
\indent
However, the use of Kinect systems as a emotional recognition tool has not been reported.
\section{Methods}
\subsection{Experiment}
\indent
$59$ healthy young subjects($32$ females and $27$ males) from University of Chinese Academy of Sciences participated in this study. They reported no injuries, illnesses or other condition influence their gait patterns.
There is a $6$ meters long footpath in the experiment environment, and at the two sides of the footpath is two Kinect camera. The experiment environment is shown in Figure 1 and Figure 2\\
\begin{figure}[htbp]
  \centering
  \includegraphics[width=12cm]{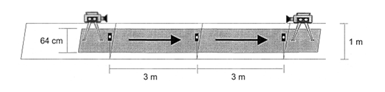}\\
  \caption{the description of the experiment environment}\label{Fig 1}
\end{figure}
\begin{figure}[htbp]
  \centering
  \includegraphics[scale=0.3]{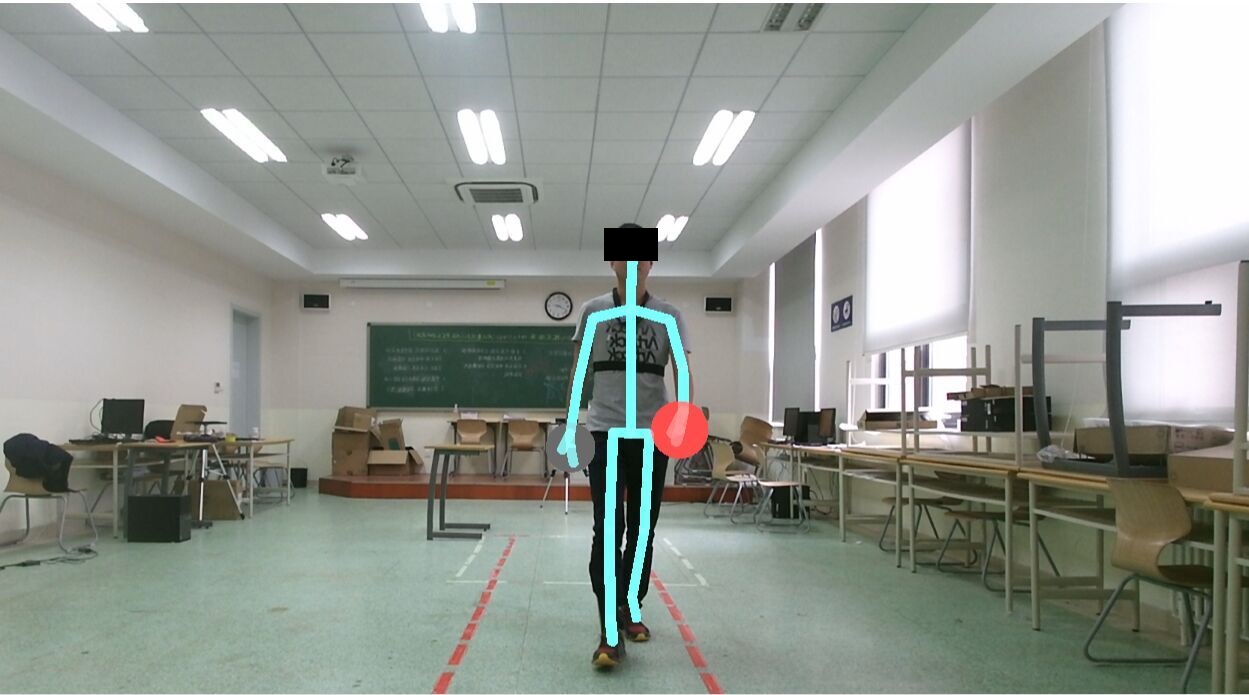}\\
  \caption{the scene of the experiment environment}\label{Fig 2}
\end{figure}
\indent
Every subject is required to walk around the footpath for 2 minutes, then they evaluate their own emotional state(angry or happy) and give a score(1 to 10). Then a stimulus video for one emotion was shown to participants, When they finish watching, they walk around the footpath for 1 minute. Every subject's gait patterns is record by two kinect camera. When they finish walking, they give two scores(1 to 10) to evaluate their own emotion state finishing watching the video and finishing the second walking.\\
\indent
The experiment is held twice. In the first round of experiment, subjects walk with natural and angry emotion. In the second round of experiment, subjects walk with natural and happy emotion.\\
\subsection{Database}
\indent
The Kinect system is placed at the two sides of the footpath, $30$Hz video data are acquired from every Kinect camera using the official Microsoft software development SDK Beta2 version and customized software(Microsoft Visual Studio 2012),One frame data contains the 3-dimensional position of $25$ joints over time, which affords a $75$ dimensional vector. The $25$ joints include head, shoulders, elbows, wrists, hands, spine ,hips, knees, ankles and feet as shown in Figure 3.\\
\begin{figure}[htbp]
  \centering
  \includegraphics[scale=.9]{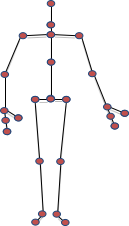}\\
  \caption{Stick figure and location of body joint centers estimated by Kinect}\label{Fig 3}
\end{figure}
\indent
The 3D coordinate system of Kinect use the Kinect camera as the origin and the unit of $3$ dimension is meter. The 3D coordinate system of Kinect is shown in Figure 4.
\begin{figure}[htbp]
  \centering
  \includegraphics[scale=.7]{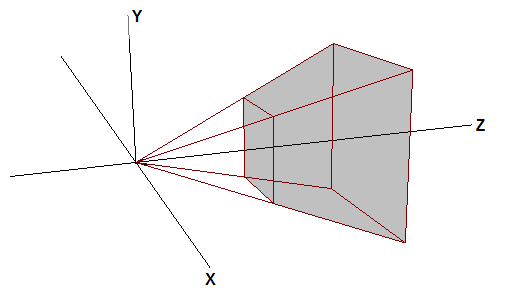}\\
  \caption{Kinect 3D coordinate}\label{Fig 4}
\end{figure}
\subsection{Data processing}
\subsubsection{Preprocessing}
\paragraph{Significant joints selecting}
According to sport anatomy theory, some joints don't change much when people walk, So we select $14$ significant joints ,which include spinebase, neck, shoulders, wrists ,elbows, hips , knees ,and ankles ,to analyze gait patterns, then one frame data contains the 3-dimmension position of $14$ significant joints, which affords a $42$ dimension vector:\\
\begin{equation}
j_t=[x_1,y_1,z_1,x_2,y_2,z_2,\dots,x_{14},y_{14},z_{14}]
\end{equation}
\indent
Suppose the process of one subject walk around the footpath one time(under an emotional state) consists of $T$ frames, called one walk. the complete dataset of one walk is described by the matrix:\\
\begin{equation}
J_t=[j_1,j_2,\dots,j_t,\dots,j_T]^T
\end{equation}
\paragraph{Coordinate system transforming}
Considering the different people have different position relative to Kinect camera when walk on the footpath, so using Kinect coordinate system has error in gait pattern analysis. To eliminate the error , we change the coordinate system by use the position of spinebase in every frame of data as the origin.\\
\indent
In the vector ,the first three columns, ,are the coordinates of spinebase joint, so the coordinate transform is given by:\\
\begin{equation}\begin{split}
&x_i^t=x_i^t-x_1^t\\
&x_i^t=x_i^t-x_1^t\\
&x_i^t=x_i^t-x_1^t\\
&(1\leq t \leq T,2 \leq i \leq 14)
\end{split}\end{equation}
\paragraph{Slider window gauss filter}
The gait data set of every walk has noises and burrs, To smooth the data set, we apply slider window gauss filter to every column of matrix $J$, the length of the window is $5$ and the convolution kernel $c=[1,4,6,4,1]/16$\\
\indent
Then the process of the filtering is given by:\\
\begin{equation}
\begin{split}
&x_i^t=[x_i^t,x_i^{t+1},x_i^{t+2},x_i^{t+3},x_i^{t+4}]\bm{\cdot} c\\
&y_i^t=[y_i^t,y_i^{t+1},y_i^{t+2},y_i^{t+3},y_i^{t+4}]\bm{\cdot} c\\
&z_i^t=[z_i^t,z_i^{t+1},z_i^{t+2},z_i^{t+3},z_i^{t+4}]\bm{\cdot} c\\
&(1 \leq t \leq T,1 \leq i \leq 14)
\end{split}
\end{equation}
\paragraph{Differential operation}
Considering the change of joints' position between each frame can reflect the people's gait patterns more well than the joints' position itself, so we apply the differential operation on the matrix $J$.\\
\indent
The differential operation is given by:
\begin{equation}
j_{t-1}=j_t-j_{t-1}(2 \leq t \leq T)
\end{equation}
\paragraph{Data segmentation}
Considering in the process of one subject walk around the footpath one time(one walk) have several straight walk segment, and the joints coordinate acquired by Kinect is not accurate when people turn round on the side of footpath. So we cut the complete dataset of one walk into several straight walk segments. Some segments record the gait patterns when subjects face to the camera, called front segments, the other segments record the gait patterns when subjects are back to the camera, called back segments. In order to ensure each segment covers at least one stride, we only choose the segment which contains at least $40$ frames to analyze.\\
\indent Suppose one walk $J$ contains $n$ front segments and $m$ back segments, which is described by a series of matrices:
\begin{equation}
\begin{cases}
&{Front}_i,1 \leq i \leq n\\
&{Back}_j,1 \leq j \leq m\\
\end{cases}
\end{equation}
\subsubsection{Feature Extraction}
\indent
The gait patterns between the front segment and the back segment is pretty different, so we extract features from front segments and back segments separately.\\
\indent
From what has been presented above, the process of one subject walk around the footpath one time(with an emotion state),called one walk, described by a matrix $J$, contains $n$ front segments and $m$ back segments.\\
\indent
First, we extract features from front segments. Human walking is periodic, and each segment covers at least one stride; therefore, Fourier transformation is applied to model the behavior of the each front segment $Front_i$,Do Fourier transformation on each column of $Front_i$,The main frequency $f_1^i,f_2^i,\dots,f_{42}^i$ and the corresponding phase $\varphi_1^i,\varphi_2^i,\dots,\varphi_{42}^i,$ are extracted.Because of one walk contains $n$ front segments and different processes of walk has different number of segments. We select mean features of every front segment. Then get a $84$ dimension vector:\\
\begin{equation}
{Feature}_{front}=\frac{1}{n}\sum_i^n[f_1^i,f_2^i,\dots,f_{42}^i,\varphi_1^i,\varphi_2^i,\dots,\varphi_{42}^i]
\end{equation}
\indent
Second, extract features from back segments in same way, get another 84 dimension vector:\\
\begin{equation}
{Feature}_{back}=\frac{1}{m}\sum_i^n[f_1^i,f_2^i,\dots,f_{42}^i,\varphi_1^i,\varphi_2^i,\dots,\varphi_{42}^i]
\end{equation}
Combine ${Feature}_{front}$ and ${Feature}_{back}$ to get a 168 dimension vector:\\
\begin{equation}
Feature=[{Feature}_{front},{Feature}_{back}]
\end{equation}
\indent
The vector $Feature$ describe the process of one subject walk on the footpath one time(with an emotional state), and the classification base on these features.
\section{Classification and Result}
\indent
For recognition, several standard classifiers are compared. NaiveBayes, Random Forests , LibSVM and SMO classifiers are used for classification. And the recognition rate is calculated using 10-fold cross validation.\\
\indent
Table 1 lists the accuracy of each classifier to recognize the natural emotion and angry emotion , the gait datasets are collected by one Kinect camera ,called KINECT1, in first round of experiment.\\
\begin{table}[htbp]
  \centering
  \caption{the accuracy of recognition of natural and angry emotion, The gait datasets are collected by KINECT1}\label{Table 1}
  \begin{tabular}{ccccc}
  \hline
  Classifier & NaiveBayes & RandomForests & LibSVM & SMO\\
  \hline
  Accuracy(\%) & \bm{$77.5862}$ &$53.4483$ &$64.6552$ &$56.0345$\\
  \hline
  \end{tabular}
\end{table}
\indent
Table 2 lists the accuracy of each classifier to recognize the natural emotion and angry emotion , the gait datasets are collected by the other Kinect camera, called KINECT2, in first round of experiment.\\
\begin{table}[htbp]
  \centering
  \caption{the accuracy of recognition of natural and angry emotion, The gait datasets are collected by KINECT2}\label{Table 2}
  \begin{tabular}{ccccc}
  \hline
  Classifier & NaiveBayes & RandomForests & LibSVM & SMO\\
  \hline
  Accuracy(\%) & \bm{$72.0339$} &$53.3898$ &\bm{$72.0339$} &$54.2373$\\
  \hline
  \end{tabular}
\end{table}
\indent
Table 3 and Table 4 list the accuracy of each classifier to recognize the natural emotion and happy emotion , the gait datasets are collected by KINECT1 and KINECT2, in second round of experiment.\\
\begin{table}[htbp]
  \centering
  \caption{the accuracy of recognition of natural and happy emotion, The gait datasets are collected by KINECT1}\label{Table 3}
  \begin{tabular}{ccccc}
  \hline
  Classifier & NaiveBayes & RandomForests & LibSVM & SMO\\
  \hline
  Accuracy(\%) & \bm{$71.1846$} & $50.8475$ & $58.4746$ & $-$\\
  \hline
  \end{tabular}
\end{table}
\begin{table}[htbp]
  \centering
  \caption{the accuracy of recognition of natural and happy emotion, The gait datasets are collected by KINECT2}\label{Table 4}
  \begin{tabular}{ccccc}
  \hline
  Classifier & NaiveBayes & RandomForests & LibSVM & SMO\\
  \hline
  Accuracy(\%) &$60.1695$ &$53.3898$ &$-$ &$-$\\
  \hline
  \end{tabular}
\end{table}
\indent
Table 5 and Table 6 list the accuracy of each classifier to recognize the angry emotion and happy emotion , the gait datasets are collected by KINECT1 and KINECT2, in first and second round of experiment.\\
\begin{table}[htbp]
  \centering
  \caption{the accuracy of recognition of angry and happy emotion, The gait datasets are collected by KINECT1}\label{Table 5}
  \begin{tabular}{ccccc}
  \hline
  Classifier & NaiveBayes & RandomForests & LibSVM & SMO\\
  \hline
  Accuracy(\%) & $55.0847$ &$54.2373$ &$62.7119$ &$54.2373$\\
  \hline
  \end{tabular}
\end{table}
\begin{table}[htbp]
  \centering
  \caption{the accuracy of recognition of angry and happy emotion, The gait datasets are collected by KINECT2}\label{Table 6}
  \begin{tabular}{ccccc}
  \hline
  Classifier & NaiveBayes & RandomForests & LibSVM & SMO\\
  \hline
  Accuracy(\%) & $-$ &$54.5423$ &$-$ &$50.0000$\\
  \hline
  \end{tabular}
\end{table}
\section{Discussion}
\indent
Recognition of emotions in gait patterns is challenge from data mining. Expression of emotions varies within subjects, ground truth of labeled data is not assured in effective computing and also gait patterns are highly individual. But the gait is capable to give cues about ones emotion state.\\
\indent
In comparison to marked-based system, Kinect may be inaccuracy in joint center estimation, it could be a possible cause for a offset error.\\
\indent
From the result we can see that the recognition rate between natural and unnatural emotion can reach more than $70\%$,But the recognition rate between different unnatural emotions(angry and happy) is not very good. It shows that people's gait may be similar with angry and happy emotional states.\\
\indent
Within significant joints selecting approach for data preprocessing,results shows good performance regarding accuracy, Using the whole 25 joints position leads to lower recognition rates, which can not reach $70\%$.
\section{Conclusion and Future Work}
\indent
This work gives a new referable method to do emotion recognition, by using Kinect system to acquire gait datasets and do feature extraction and classification. The natural and unnatural emotions are in general better recognizable in gait patterns, but the recognition of angry and happy emotions is not very good. And due to low number of subjects, the recognition rate for emotions in walking remains around chance level in. So, further improvement is required. This includes more gait patterns data acquisition by recruit more subjects and further improvement of feature extraction as well as classification.
\section{Acknowledgement}
\indent
The authors would like to thank Prof.Tingshao Zhu and Prof.Baobin Li for their advice and useful discussion.
\section*{References}

\bibliography{KinectPaperReferences}

\begin{thebibliography}{10}
\expandafter\ifx\csname url\endcsname\relax
  \def\url#1{\texttt{#1}}\fi
\expandafter\ifx\csname urlprefix\endcsname\relax\def\urlprefix{URL }\fi
\expandafter\ifx\csname href\endcsname\relax
  \def\href#1#2{#2} \def\path#1{#1}\fi

\bibitem{bealeaffect}
C.~P.~R. Beale, Affect and emotion in human-computer interaction.

\bibitem{pollick2001perceiving}
F.~E. Pollick, H.~M. Paterson, A.~Bruderlin, A.~J. Sanford, Perceiving affect
  from arm movement, Cognition 82~(2) (2001) B51--B61.

\bibitem{1323098}
A.~Kale, A.~Sundaresan, A.~Rajagopalan, N.~Cuntoor, A.~Roy-Chowdhury,
  V.~Kruger, R.~Chellappa, Identification of humans using gait, Image
  Processing, IEEE Transactions on 13~(9) (2004) 1163--1173.
\newblock \href {http://dx.doi.org/10.1109/TIP.2004.832865}
  {\path{doi:10.1109/TIP.2004.832865}}.

\bibitem{Gaukrodger2013}
S.~Gaukrodger, A.~Peruzzi, G.~Paolini, A.~Cereatti, S.~Cupit, J.~Hausdorff,
  A.~Mirelman, C.~U. Della, M38, Gait \& posture 37~(Supplement 1) (2013) S31.
\newblock \href {http://dx.doi.org/10.1016/j.gaitpost.2012.12.062}
  {\path{doi:10.1016/j.gaitpost.2012.12.062}}.

\bibitem{Pogrzeba2013}
Pogrzeba, Loreen, Wacker, Markus, Jung, Bernhard, P22, Gait \& Posture
  38~(Supplement 1) (2013) S94--S95.
\newblock \href {http://dx.doi.org/10.1016/j.gaitpost.2013.07.195}
  {\path{doi:10.1016/j.gaitpost.2013.07.195}}.

\bibitem{Stone201557}
E.~Stone, M.~Skubic, M.~Rantz, C.~Abbott, S.~Miller,
  \href{http://www.sciencedirect.com/science/article/pii/S0966636214006833}{Average
  in-home gait speed: Investigation of a new metric for mobility and fall risk
  assessment of elders}, Gait \& Posture 41~(1) (2015) 57 -- 62.
\newblock \href
  {http://dx.doi.org/http://dx.doi.org/10.1016/j.gaitpost.2014.08.019}
  {\path{doi:http://dx.doi.org/10.1016/j.gaitpost.2014.08.019}}.
\newline\urlprefix\url{http://www.sciencedirect.com/science/article/pii/S0966636214006833}

\bibitem{7}
J.~Montepare, S.~Goldstein, A.~Clausen,
  \href{http://dx.doi.org/10.1007/BF00999605}{The identification of emotions
  from gait information}, Journal of Nonverbal Behavior 11~(1) (1987) 33--42.
\newblock \href {http://dx.doi.org/10.1007/BF00999605}
  {\path{doi:10.1007/BF00999605}}.
\newline\urlprefix\url{http://dx.doi.org/10.1007/BF00999605}

\bibitem{8}
D.~Janssen, W.~Schöllhorn, J.~Lubienetzki, K.~Fölling, H.~Kokenge, K.~Davids,
  \href{http://dx.doi.org/10.1007/s10919-007-0045-3}{Recognition of emotions in
  gait patterns by means of artificial neural nets}, Journal of Nonverbal
  Behavior 32~(2) (2008) 79--92.
\newblock \href {http://dx.doi.org/10.1007/s10919-007-0045-3}
  {\path{doi:10.1007/s10919-007-0045-3}}.
\newline\urlprefix\url{http://dx.doi.org/10.1007/s10919-007-0045-3}

\bibitem{karg2009two}
M.~Karg, R.~Jenke, K.~K{\"u}hnlenz, M.~Buss, A two-fold pca-approach for
  inter-individual recognition of emotions in natural walking., in: MLDM
  Posters, 2009, pp. 51--61.

\bibitem{5349438}
M.~Karg, R.~Jenke, W.~Seiberl, K.~Kuuhnlenz, A.~Schwirtz, M.~Buss, A comparison
  of pca, kpca and lda for feature extraction to recognize affect in gait
  kinematics, in: Affective Computing and Intelligent Interaction and
  Workshops, 2009. ACII 2009. 3rd International Conference on, 2009, pp. 1--6.
\newblock \href {http://dx.doi.org/10.1109/ACII.2009.5349438}
  {\path{doi:10.1109/ACII.2009.5349438}}.

\bibitem{5439949}
M.~Karg, K.~Kuhnlenz, M.~Buss, Recognition of affect based on gait patterns,
  Systems, Man, and Cybernetics, Part B: Cybernetics, IEEE Transactions on
  40~(4) (2010) 1050--1061.
\newblock \href {http://dx.doi.org/10.1109/TSMCB.2010.2044040}
  {\path{doi:10.1109/TSMCB.2010.2044040}}.

\bibitem{Wang20111780}
J.~Wang, R.~Chen, X.~Sun, M.~F. She, Y.~Wu,
  \href{http://www.sciencedirect.com/science/article/pii/S1877705811018327}{Recognizing
  human daily activities from accelerometer signal}, Procedia Engineering 15
  (2011) 1780 -- 1786, \{CEIS\} 2011.
\newblock \href
  {http://dx.doi.org/http://dx.doi.org/10.1016/j.proeng.2011.08.331}
  {\path{doi:http://dx.doi.org/10.1016/j.proeng.2011.08.331}}.
\newline\urlprefix\url{http://www.sciencedirect.com/science/article/pii/S1877705811018327}

\bibitem{Clark20131064}
R.~A. Clark, Y.-H. Pua, A.~L. Bryant, M.~A. Hunt,
  \href{http://www.sciencedirect.com/science/article/pii/S0966636213001884}{Validity
  of the microsoft kinect for providing lateral trunk lean feedback during gait
  retraining}, Gait \& Posture 38~(4) (2013) 1064 -- 1066.
\newblock \href
  {http://dx.doi.org/http://dx.doi.org/10.1016/j.gaitpost.2013.03.029}
  {\path{doi:http://dx.doi.org/10.1016/j.gaitpost.2013.03.029}}.
\newline\urlprefix\url{http://www.sciencedirect.com/science/article/pii/S0966636213001884}

\bibitem{6096866}
F.~Kondori, S.~Yousefi, H.~Li, S.~Sonning, S.~Sonning, 3d head pose estimation
  using the kinect, in: Wireless Communications and Signal Processing (WCSP),
  2011 International Conference on, 2011, pp. 1--4.
\newblock \href {http://dx.doi.org/10.1109/WCSP.2011.6096866}
  {\path{doi:10.1109/WCSP.2011.6096866}}.

\bibitem{6338001}
A.~Fern'ndez-Baena, A.~Susin, X.~Lligadas, Biomechanical validation of
  upper-body and lower-body joint movements of kinect motion capture data for
  rehabilitation treatments, in: Intelligent Networking and Collaborative
  Systems (INCoS), 2012 4th International Conference on, 2012, pp. 656--661.
\newblock \href {http://dx.doi.org/10.1109/iNCoS.2012.66}
  {\path{doi:10.1109/iNCoS.2012.66}}.

\bibitem{6090521}
B.~Lange, C.-Y. Chang, E.~Suma, B.~Newman, A.~Rizzo, M.~Bolas, Development and
  evaluation of low cost game-based balance rehabilitation tool using the
  microsoft kinect sensor, in: Engineering in Medicine and Biology Society,
  EMBC, 2011 Annual International Conference of the IEEE, 2011, pp. 1831--1834.
\newblock \href {http://dx.doi.org/10.1109/IEMBS.2011.6090521}
  {\path{doi:10.1109/IEMBS.2011.6090521}}.

\bibitem{6165146}
J.~Tong, J.~Zhou, L.~Liu, Z.~Pan, H.~Yan, Scanning 3d full human bodies using
  kinects, Visualization and Computer Graphics, IEEE Transactions on 18~(4)
  (2012) 643--650.
\newblock \href {http://dx.doi.org/10.1109/TVCG.2012.56}
  {\path{doi:10.1109/TVCG.2012.56}}.

\bibitem{Auvinet2015722}
E.~Auvinet, F.~Multon, C.-E. Aubin, J.~Meunier, M.~Raison,
  \href{http://www.sciencedirect.com/science/article/pii/S0966636214006705}{Detection
  of gait cycles in treadmill walking using a kinect}, Gait \& Posture 41~(2)
  (2015) 722 -- 725.
\newblock \href
  {http://dx.doi.org/http://dx.doi.org/10.1016/j.gaitpost.2014.08.006}
  {\path{doi:http://dx.doi.org/10.1016/j.gaitpost.2014.08.006}}.
\newline\urlprefix\url{http://www.sciencedirect.com/science/article/pii/S0966636214006705}

\bibitem{Yeung2014532}
L.~Yeung, K.~C. Cheng, C.~Fong, W.~C. Lee, K.-Y. Tong,
  \href{http://www.sciencedirect.com/science/article/pii/S0966636214006122}{Evaluation
  of the microsoft kinect as a clinical assessment tool of body sway}, Gait \&
  Posture 40~(4) (2014) 532 -- 538.
\newblock \href
  {http://dx.doi.org/http://dx.doi.org/10.1016/j.gaitpost.2014.06.012}
  {\path{doi:http://dx.doi.org/10.1016/j.gaitpost.2014.06.012}}.
\newline\urlprefix\url{http://www.sciencedirect.com/science/article/pii/S0966636214006122}

\bibitem{Galna20141062}
B.~Galna, G.~Barry, D.~Jackson, D.~Mhiripiri, P.~Olivier, L.~Rochester,
  \href{http://www.sciencedirect.com/science/article/pii/S0966636214000241}{Accuracy
  of the microsoft kinect sensor for measuring movement in people with
  parkinson's disease}, Gait \& Posture 39~(4) (2014) 1062 -- 1068.
\newblock \href
  {http://dx.doi.org/http://dx.doi.org/10.1016/j.gaitpost.2014.01.008}
  {\path{doi:http://dx.doi.org/10.1016/j.gaitpost.2014.01.008}}.
\newline\urlprefix\url{http://www.sciencedirect.com/science/article/pii/S0966636214000241}

\bibitem{khoshelham2012accuracy}
K.~Khoshelham, S.~O. Elberink, Accuracy and resolution of kinect depth data for
  indoor mapping applications, Sensors 12~(2) (2012) 1437--1454.

\bibitem{stone2011evaluation}
E.~Stone, M.~Skubic, Evaluation of an inexpensive depth camera for in-home gait
  assessment, Journal of Ambient Intelligence and Smart Environments 3~(4)
  (2011) 349--361.

\end{thebibliography}

\end{document}